# Assessment of Low Scaling Approximations to EOM-CCSD Method for Ionization Potential

*Achintya Kumar Dutta[a,b]\*, Nayana Vaval[a] and Sourav Pal[c]\**

[a]*Physical Chemistry Division, CSIR-National Chemical Laboratory, Pune-411008, India*

[b]*Present address : Max plank Institute for Chemical Energy Conversion, Mulheim an Der Ruhr, Germany*

[c]*Department of Chemistry, Indian Institute of Technology Bombay, Powai, Mumbai 400 076, India*

In this paper, we investigate the performance of different approximate variants of the EOM-CCSD method for calculation of ionization potential, as compared to EOM-CCSDT reference values. The errors in the various approximate variants of the EOM-CCSD method are quite different for different kind of ionized states. None of the approximate variants including the original EOM-CCSD method gives a uniform performance over outer valence, inner valence, and core ionization, favoring one or other, depending upon nature of the approximation used.

\*achintya-kumar.dutta@cec.mpg.de

\*spal@chem.iitb.ac.in

# 1. Introduction:

Ionization energy is one the intrinsic properties of atoms and molecules, which has continued to fascinate generations of experimentalists and theoreticians[1]. The accurate determination of ionization energy is of extreme importance in biology and chemistry. In spite of the tremendous advancement of spectroscopic techniques in recent times, experimental determination of ionization energies is often troublesome. Therefore, theoretical calculations are generally utilized as supportive, sometimes the sole mean for understanding of electron detachment induced phenomenon.

The various theoretical methods available for the calculations of ionization potential (IP) are broadly classified into two categories[2]. The first one is the so-called Δ based techniques, where two separate calculations are required for the ion and the neutral species and the IP is obtained as the difference of energies obtained in two separate calculations. The second strategy consists of the so-called 'direct difference of energy' scheme, which describes ionization as a transition process between the neutral molecule and the ion. The coupled cluster linear response theory[3,4], quasi-degenerate perturbation theories, and Green function based methods[5-8] fall into this second category and can be unified under the general framework of equation of motion (EOM) approach[9]. The direct difference of energy scheme has some significant advantage over the Δ-based technique. Firstly, the direct difference of energy approach generates the IP in a single calculation, not as the difference of two big numbers as in Δ technique. Secondly, it gives information about the transition process and transition probability, which allows the simulation of experimental spectroscopic signatures.

Among the different variants of EOM methods available, the equation of motion coupled cluster (EOM-CC) method[10-13] provides the most systematic way of balanced inclusion of dynamic and non-dynamic correlations. The EOM-CC approach for the ionization problem[14] (EOMIP-CC) is generally used in singles and doubles truncation (EOMIP-CCSD) and provides an easy way to (0,1) sector of Fock space[15], without venturing into the conceptual difficulties of the corresponding multi-reference coupled cluster (FSMRCC) method[16-18]. The EOMIP-CCSD method scales as $N^6$ power of the basis set and has similar storage requirement as that of the single-reference coupled cluster method, which prohibits its use beyond medium sized molecules in a reasonable basis set.

The coupled cluster method shares an intriguing relationship with many-body perturbation theory[19]. So, the most obvious way of deriving an approximation to the coupled cluster method is based on perturbation orders. Nooijen and Sniders were the first to propose the use of MBPT(2) amplitudes in place of coupled cluster ansatz in the context of IP calculations[20]. Stanton and Gauss[21] generalized this idea to define a hierarchy of approximations to the standard EOM-CCSD method called EOM-CCSD(n), where the reference state energy is complete up to $n^{th}$ order in perturbation. The method is size-extensive for each value of n and the lowest order of approximation to it leads to EOM-CCSD(2) method with MBPT(1) ground state wave function and MBPT(2) ground state energy. Similar ideas were persuaded by Bartlett and co-workers in the context of excitation energy[22], and Dutta *et. al.* for electron affinity[23] and spin-flip[24] variants of EOM-CC.

For ionization problem, the EOM-CCSD(2) approximation offers a significant reduction in computational requirements[21]. The EOMIP-CCSD(2) method has an iterative $N^5$ scaling and does not involve (ab|cd) integral, leading to a drastic reduction in the storage requirements. Therefore, it can be applied to very large systems. Pal and co-workers[25] have demonstrated that the EOMIP-CCSD(2) method can be used to calculate the structure and vibrational frequency of doublet radicals with accuracy similar to that of the standard EOMIP-CCSD method. It is possible to further approximate the EOMIP-CCSD(2) method, by using a diagonal approximation of the doubles-doubles block. The idea is similar to the P-EOM-MBPT(2) method developed for EE[22] and EA[23]. A more accurate version of EOMIP-CCSD(2) version is developed by Pal and co-workers[26] which gives results comparable to the standard EOMIP-CCSD for both IP and geometry. Krylov and co-workers[27] have developed the IP-CISD method for calculation of IP, which is also iterative $N^5$ scaling and does not require four particle intermediates. Recently, Schütz and co-workers[28] have implemented a linear scaling CC2 method for ionization potential.

In spite of the availability of the various approximate variants of EOMIP-CC, no benchmark results are available with their relative accuracy for IP calculation. Therefore, a systematic study to investigate the source and the magnitude of the error is absolutely necessary before proceeding with the routine use of the approximate EOMIP-CCSD method. Some nice benchmark studies on the *ab-initio* calculation of Ionization potential has been reported in recent times. Sheril and co-workers[29] have constructed a benchmark for 24 organic dyes, using extrapolated CCSD(T) results were used as the reference. However, the reported ionization potentials were only restricted to ionization from highest occupied molecular orbitals (HOMO). Now, it is well known that the Koopmans' picture breaks down for inner valence orbitals and consequently ionization from inner valence orbitals behave very differently than the outer valence orbitals. Recently, Ortiz and co-workers[30] have published a benchmark of SAC-CI results, where the values were compared against experimental values. Now, it is not always straightforward to assign experimental results with theoretical calculations and different experiments can lead to different interpretation of the measured ionization potential[31]. For example, detachment energy obtained from optical spectroscopy or Multi-Photon Ionization and Resonance Ionization Mass Spectrometry corresponds to adiabatic ionization energy. On the other hand, photoelectron spectroscopy experiments result in vertical ionization energy. Incompleteness in one-electron basis function can be another major source of error while comparing with experiments. Therefore, in our opinion, it is justified to use the same basis set in a higher-level method as the reference to get an unambiguous estimation of the error due to approximation in the wave-function.

The aim of this paper is to perform a benchmark study for outer valence, inner valence and core IP in different approximate variants of the EOMIP-CCSD method against the EOMIP-CCSDT method, and to rationalize the source of error in the former.

The paper is organized as follows. Section 2 gives a brief overview on the different approximate variants of the EOMIP-CCSD method and computational details of the calculations. The trends in the numerical results and sources of errors are analyzed in section 3. Section 4 gives the concluding remarks.

## 2. Theory and Computational Details

In the EOM framework, the k[th] excited state is generated from a reference state[9] by action of a linear excitation operator $\hat{R}_k$

$$|\psi_k\rangle = \hat{R}_k |\psi_0\rangle \qquad (1)$$

The explicit form of $\hat{R}_k$ depends upon the nature of the excited state. For ionized state,

$$\hat{R}_k^{IP} = \sum_i R_i(k) i + \sum_{i>j,a} R_{ij}^a(k) \hat{a}^\dagger \hat{j} \hat{i} + \ldots\ldots\ldots \qquad (2)$$

This is general EOM framework. The coupled cluster enters into the picture[10] with the fact that in EOM-CC the correlated reference state is generated from a single Slater determinant by the action of an exponential operator as following

$$|\psi_0\rangle = e^T |\phi_0\rangle \qquad (3)$$

$|\phi_0\rangle$ is generally, but not necessarily, a Hartree-Fock determinant and T=$T_1$+$T_2$+$T_3$+…… $T_n$, where

$$\hat{T}_1 = \sum_{ia} t_i^a \{a_a^\dagger a_i\}$$

$$\hat{T}_2 = \frac{1}{4} \sum_{ijab} t_{ij}^{ab} \{a_a^\dagger a_b^\dagger a_j a_i\}, \qquad (4)$$

$$\hat{T}_3 = \frac{1}{36} \sum_{ijkabc} t_{ijk}^{abc} \{a_a^\dagger a_b^\dagger a_c^\dagger a_k a_j a_i\}$$

,

These amplitudes are generally obtained by the iterative solution of a system of coupled nonlinear equations.

In the EOMIP-CC framework, the final ionized states are obtained by diagonalising the coupled cluster similarity transformed Hamiltonian within (N-1) electron space[14].

$$\bar{H} = e^{-T}He^{T} = \left(He^{T}\right)_{c} \quad (5)$$

The resulting IP values for the principle peaks are identical to that obtained from the solution of the (0,1) sector of the Fock space multi reference coupled cluster (FSMRCC) method[15].

The similarity transformed Hamiltonian can be written more explicitly as

$$\bar{H} = \begin{pmatrix} \langle f_0|\bar{H}|f_0\rangle & \langle f_0|\bar{H}|S\rangle & \langle f_0|\bar{H}|D\rangle \\ 0 & \langle S|\bar{H}|S\rangle & \langle S|\bar{H}|D\rangle \\ 0 & \langle D|\bar{H}|S\rangle & \langle D|\bar{H}|D\rangle \end{pmatrix} = \begin{pmatrix} \bar{H}_{00} & \bar{H}_{OS} & \bar{H}_{0D} \\ 0 & \bar{H}_{SS} & \bar{H}_{SD} \\ 0 & \bar{H}_{DS} & \bar{H}_{DD} \end{pmatrix} \quad (6)$$

$|S\rangle$ and $|D\rangle$ in the above equation represents 1h and 2h1p determinant, which is generated from the Hartree-Fock wave-function as

$$|S\rangle = \tau_1|\phi_0\rangle \quad (7)$$

$$|D\rangle = \tau_2|\phi_0\rangle$$

where

$$\tau_1 = \sum_i \tau^i i \quad (8)$$

$$\tau_2 = \sum_{b,j>i} \tau_b^{ji} b^\dagger$$

The EOM step in the EOM-CCSD scales as iterative $O(N^5)$, however, the solution of coupled cluster scales as iterative $O(N^6)$. It should be noted that 4 particle intermediates are absent in the EOM part.

In IP-CISD method of Krylov and coworker[27], one diagonalizes $H$ instead of $\bar{H}$

$$H = \begin{pmatrix} \langle f_0|\hat{H}|f_0\rangle & 0 & \langle f_0|\hat{H}|D\rangle \\ 0 & \langle S|\hat{H}|S\rangle & \langle S|\hat{H}|D\rangle \\ 0 & \langle D|\hat{H}|S\rangle & \langle D|\hat{H}|D\rangle \end{pmatrix} = \begin{pmatrix} H_{00} & 0 & H_{0D} \\ 0 & H_{SS} & H_{SD} \\ 0 & H_{DS} & H_{DD} \end{pmatrix} \quad (9)$$

IP-CISD scales iterative $O(N^5)$ and has much less storage requirement as it does not require 3 and 4 particle intermediate.

An alternative way of reducing computational cost is to truncate $\bar{H}$ using perturbation theory. The coupled cluster method has an intriguing relationship with MBPT method and various orders of MBPT can be obtained from the suitable lower order iterations of coupled cluster equations[19]. For example, the lowest order approximation to CCSD leads to the MBPT(2) method. Therefore, a natural way of truncating the CCSD similarity transformed Hamiltonian should be based on perturbative orders. Two slightly different perturbative approaches have been reported in the literature[20,21]. Nooijen and Snijders[20] first proposed a truncation of the CCSD effective Hamiltonian based on the perturbation order. Stanton and Gauss[21] generalized the idea within the EOM framework. They have performed a perturbational expansion of the effective Hamiltonian.

$$\bar{H} = \left(He^T\right)_c = \bar{H}^{[1]} + \bar{H}^{[2]} + \bar{H}^{[3]} + \ldots\ldots + \bar{H}^{[n]} \quad (10)$$

The subscript c in the above equation represents the connectedness of $T$ with $H$ and bracketed superscript represents the order in perturbation and. Equation (10) leads to a set of hierarchical approximation to the full $\bar{H}$, and the diagonalization of the modified similarity transformed Hamiltonian offers a set of hierarchical approximation to the corresponding EOM-CC final states, denoted as EOMCCSD(n). The similarity transformed Hamiltonian truncated at any arbitrary $n^{th}$ order, only contains terms up to $n^{th}$ order in perturbation, which makes the method size extensive of for all values of n. At large value of n, the $\bar{H}^{[2]}$ converges to the full $\bar{H}$ and consequently the EOM-CCSD(n) leads to the standard EOM-CCSD method. Truncation at second order (n=2), leads to EOM-CCSD(2), with an MBPT(1) reference state wave function and MBPT(2) reference energy. Consequently, the CC amplitudes in EOM-CCSD(2) can be replaced by the MBPT(2) amplitudes

$$\begin{aligned}\bar{H} &= \left(H e^{T}\right)_{c} \\ &\approx \left(H e^{T'}\right)_{c}\end{aligned} \quad (11)$$

Where the second order perturbative approximation to the T amplitudes ($T'$) can be expressed as

$$\begin{aligned}T_1' &= \frac{f_{ia}}{\varepsilon_i - \varepsilon_a} \\ T_2' &= \frac{\langle ab||ij\rangle}{\varepsilon_i + \varepsilon_j - \varepsilon_a - \varepsilon_b}\end{aligned} \quad (12)$$

$T_1'$ is zero for RHF and UHF MBPT(2) reference. One can obtain a modified similarity transformed Hamiltonian $\bar{H}^{[2]}$ using these $T'$ amplitudes, which can be diagonalized to get ionization energy.

$$\bar{H}^{[2]} = \begin{pmatrix} \langle\phi_0|\bar{H}^{[2]}|\phi_0\rangle & 0 & \langle\phi_0|\bar{H}^{[2]}|D\rangle \\ 0 & \langle S|\bar{H}^{[2]}|S\rangle & \langle S|\bar{H}^{[2]}|D\rangle \\ 0 & \langle D|\bar{H}^{[2]}|S\rangle & \langle D|\bar{H}^{[2]}|D\rangle \end{pmatrix} = \begin{pmatrix} \bar{H}^{[2]}_{00} & 0 & \bar{H}^{[2]}_{0D} \\ 0 & \bar{H}^{[2]}_{SS} & \bar{H}^{[2]}_{SD} \\ 0 & \bar{H}^{[2]}_{DS} & \bar{H}^{[2]}_{DD} \end{pmatrix} \quad (13)$$

This approach is slightly different from that originally proposed by Nooijen and Snijders[20]. Reference 21 should be consulted for an elaborate discussion on the differences between these two approaches.

In this approach, the reference state energy reduces to the MBPT(2) ground state energy, with the accompanying reduction in the computational scaling from iterative $N^6$ to non-iterative $N^5$ for the reference state. There are still some terms of $\bar{H}^{[2]}$ that scale as $N^6$, however, scaling of those steps can be reduced to iterative $N^5$ by calculating them on the fly. Moreover, the truncation at MBPT(2) ensures the absence of 4 particle intermediates, which anyways remain absent from the EOM part in IP calculations. Therefore, the EOMIP-CCSD(2) method gives significant saving in terms of scaling, as well as storage requirement. Although, the storage requirement is slightly higher than the IP-CISD method of Krylov and co-worker as it needs the 3 particle integrals[27].

Two closely related approaches to EOMIP-CCSD(2) are partitioned EOMIP-CCSD(2) (P-EOMIP-CCSD(2) and EOMIP-CCSD(2)* method[26]. In P-EOM-CCSD(2) approach the doubles-doubles block of the $\bar{H}^{[2]}$ matrix is approximated as diagonal. It has been observed that partitioned version of EOM-CCSD(2) method provides an improvement in results compared to the standard EOM-CCSD(2) method for both EA[23] and EE[22]. Therefore, it would be interesting to extend the idea to IP problem. Here, it should be noted that partitioning approach does not provide any significant decrease in storage requirement in EOMIP-CCSD(2) method, unlike in the case of electron affinity problem[23], where it leads to a drastic decrease in the storage requirements and the method still scales as $O(N^5)$.

The lack of ground state singles amplitudes leads to decrease in orbital relaxation effect. To counter that partial triples[32,33] are included in EOMIP-CCSD(2) to account for the relaxation effect and the resulting EOMIP-CCSD(2)* method[26] gives the ionization potential as

$$\langle L_p | \bar{H}_{eff} | R_p \rangle = E_{EOMIP-CCSD(2)} + \Delta W = E_{EOMIP-CCSD(2)} + \langle \partial L | D | \partial R \rangle \qquad (14)$$

where

$$D_{ab}^{ijk} l_{ab}^{ijk} = P(ijk) l^k \langle ab \| ij \rangle - P(ijk) \sum_e l_e^{ij} \langle ek \| ab \rangle - P(ab) P(ijk) \sum_m l_a^{mk} \langle ij \| mb \rangle$$
$$D_{ijk}^{ab} r_{ijk}^{ab} = -P(ijk) \sum_e r_{ij}^e \langle ab \| ek \rangle - P(ab) P(ijk) \sum_m r_{mk}^a \langle mb \| ij \rangle$$
$$- P(ab) P(ijk) \sum_{me} r_m t_{ij}^{ae} \langle mb \| ke \rangle + P(kji) \sum_{mn} r_m t_{in}^{ab} \langle mn \| kj \rangle$$

(15)

and

$$D^{ijk}_{ab} = D^{ab}_{ijk} = W_0 - E_0 + f_{ii} + f_{jj} + f_{kk} - f_{aa} - f_{bb} \qquad (16)$$

The method scales as overall non-iterative $N^6$ but has lesser prefactor than a single ground state CCSD iteration[26].

In EOM-CC2 method, one diagonalizes the matrix

$$A = \begin{pmatrix} \langle S|\bar{H}|S\rangle & \langle S|\ddot{H}|D\rangle \\ \langle D|\ddot{H}|S\rangle & \langle D|\hat{H}|D\rangle \end{pmatrix} \qquad (17)$$

where $\ddot{H}$ is the singles transformed Hamiltonian

$$\ddot{H} = e^{-\hat{T}_1}\hat{H}e^{\hat{T}_1} \qquad (18)$$

The $\hat{T}_1$ and $\hat{T}_2$ used in $\bar{H}$ comes from a ground state CC2 equation and the doubles doubles block is diagonal same as that in P-EOM-CCSD(2). The method has a slightly higher storage requirement as the 4 virtual integrals are necessary for the ground state calculation. Table I presents scaling and storage requirements of different approximate variants of EOM-CC.

All the EOM-CCSDT, EOM-CCSD, EOM-CCSD(2), EOM-CC2 and EOMIP-CCSD(2)* calculations were performed using CFOUR[34]. The IP-CISD and P-EOMIP-CCSD(2) calculations were performed using our in-house coupled cluster codes. Experimental geometries are used for all the molecules. cc-pVTZ basis set[35] has been used for valence IP and cc-pCVDZ basis set[36] has been used for core IP. Cartesian coordinates of the studied molecules and calculated IP values are provided in the supporting information.

**3 Results and Discussion**

The physics of ionization from different occupied orbitals can be considerably different. For example, ionization from outer valence orbitals is mostly dominated 1h block of the Hamiltonian and Koopmans picture gives a very good zeroth order description of the ionization process. On the other hand for inner valence, the coupling between 1h and 2h1p blocks becomes quite important and Koopmans picture starts to breakdown[5]. For core-ionization, even the presence of the 2h1p block of the Hamiltonian is not enough to provide sufficient relaxation effect, and one needs triples correction to get sufficient accuracy. Therefore, in this study we analyze the accuracy of different approximate variants of EOM-CCSD separately for ionization outer valence, inner valence and core orbital. Our test set consist of 18 molecules consisting of $N_2$, $H_2O$, $ClF$, $H_2CO$, $CO$, $C_2H_2$, $C_2H_4$, $O_3$, $NH_3$, $F_2$, $CO_2$, $SO_2$, $N_2O$, $BN$, $HF$, $S_2$, $P_2$, and $OH^-$. The main reason for using diatomic and triatomic molecules is to keep the EOM-CCSDT calculations feasible. We have confined our analysis to the so-called 'principle peaks' i.e. the peaks dominated by 1h terms. The satellite peaks which are dominated by 2h1p terms are not considered in this study, as the EOM-CCSD method itself does not give accurate results for that kind of states.

### 3.1 Outer Valence Ionization

The ionization from outer valence orbitals is generally most easy to simulate. Especially ionization from the HOMO, which can be easily simulated by using so-called "Δ" based method.  One can use Δ CCSD(T) results as an alternative benchmark, in addition to the EOM-CCSDT method. It can be seen that both the method gives almost identical results.  A plot of IP values in Δ CCSD(T) vs EOM-CCSDT method gives an almost linear plot. A straight line fit through the data gives a slope of ≈ 1 and intercept of 0.004, which shows the similarity of the IP values in both the method used as the benchmark. Among the various approximate methods tested in this study, EOM-CCSD gives the best agreement with ΔCCSD(T) data (see Table 1), with maximum absolute error (Max. Abs. Error) of 0.26 eV, mean absolute error (MAE) of 0.09 eV and root mean square deviation (RMSD) of 0.11 eV. The ΔCCSD method shows a slightly inferior performance with Max. Abs. Error of 0.29 eV, MAE of 0.13 eV and RMSD error of 0.15 eV. The EOMIP-CCSD(2)* method gives very similar performance as that of the ΔCCSD method, with Max. Abs. error of 0.21 eV, MAE of 0.12 eV and RMSD error of 0.13 eV, although it has smaller computational requirements than that of ΔCCSD method.  It can be seen from Figure 2 that, EOM-CCSD tends to slightly overestimate the IP values as compared to Δ CCSD(T) method, whereas ΔCCSD and EOMIP-CCSD(2)* method underestimate IP values.  The original EOMIP-CCSD(2) method, on the other hand, tend to overestimate the IP values,  with  Max. Abs. Error of 0.44 eV, MAE of 0.20 eV and RMSD error of 0.26 eV. P-EOM-CCSD(2) method gives slightly inferior results as compared to EOM-CCSD(2) method, with error spread on both sides of the origin making it more unpredictable than original EOM-CCSD(2) approximation. The IP-CISD method tends to drastically underestimate the IP values with Max. Abs. error of 3.93 eV. The MAE and RMSD values are also very high at 2.59 eV and 2.64 eV The mean signed deviation at -2.59 eV, is identical in magnitude with that of the MAE value indicating the IP values are systematically underestimated in IP-CISD. This is due to the fact that in IP-CISD, the reference state is treated in Hartree-Fock level and the target ionized states are treated at CISD level. The resulting unbalance leads to more lowering of target state energy than that in the reference state energy. Consequently, IP values are heavily underestimated in IP-CISD. Although, It should be kept in mind that the IP-CISD was not originally designed for calculation of IP. The main idea behind IP-CISD approximation was to get good properties and potential energy surface at a low computational cost, for which IP-CISD has shown to do a remarkably good job[12]. It is surprising to note that the more sophisticated EOM-CC2, which gives quite accurate excitation energies[37], fails significantly in the case of IP, with a high Max. Abs. Error of 1.02 eV, MAE of 0.40 eV and RMSD error of 0.48 eV. The IP values in EOM-CC2 are significantly underestimated as compared to CCSD(T). Recently, Szalay and co-workers[38] have shown that although CC2 gives quite accurate results for valence excited states due to error cancellation, its performance deteriorates in case of Rydberg states, where the electrons are excited into very diffused virtual orbitals near the continuum and balance between ground and excited states is lost to a large extent in CC2 method. This also results in loss of accuracy in the CC2 method for IP states, which are very similar in nature to Rydberg excited states. The magnitude and distribution of error in all the approximate methods as compared to EOM-CCSDT method remains almost same as that compared to Δ CCSD(T) (see Figure 3), with a slight increase in error in Δ CCSD method (see Table 3).

It is well known that in some of the cases ionization from HOMO does not give the lowest energy IP and the effect of electron correlation is significant even for outer valence IP. In Table 4 we have analyzed the error in ionization from outer valence orbital as compared to benchmark EOM-CCSDT results. Total 64 IP states are used in the statistical analysis. The trends are a generalization of what observed in the case of ionization from HOMO. The EOM-CCSD method gives the best performance with Max. Abs. error of 0.59 eV, MAE of 0.11 eV and RMSD of 0.15 eV. The EOM-CCSD(2) method shows a Max. Abs. Error of 0.86 eV, MAE of 0.21 eV and RMSD of 0.29 eV. In both the cases, the Max. Abs. Error increases from that observed in case of ionization from HOMO. Both EOM-CCSD and EOM-CCSD(2) overestimate the IP as compared to EOM-CCSDT value. From Figure 4, it can be seen that the overestimation is more prominent in the case of EOM-CCSD(2), with a considerable higher spread of error as compared to EOM-CCSD. The P-EOM-CCSD(2) method shows quite similar performance as that of EOM-CCSD(2), with a Max. Abs. error of 0.65 eV, MAE of 0.23 eV and RMSD of 0.29 eV. The errors are also more evenly distributed around the origin, however, the spread of the error is more than that in EOM-CCSD(2). The IP-CISD method continues to show inferior performance with MSD value of -2.60 eV, which essentially means that the results are always heavily underestimated in IP-CISD method. The EOMIP-CCSD(2)* method gives very similar performance as that of the EOM-CCSD method, with slightly higher Max. Abs. error of 0.87 eV. However, the results in EOMIP-CCSD(2)* method tends to be underestimated as opposed to be overestimated in EOM-CCSD. The EOM-CC2 method, on the other hand, gives a quite inferior performance for outer valence IP with a Max. Abs. error of 1.03 eV, MAE of 0.44 eV and RMSD of 0.52 eV. The IP values in EOM-CC2 are generally underestimated, with a considerable spread of errors.

## 3.2 Inner Valence Ionization

The ionization from inner valence orbitals is much more complicated than that in outer valence. The simple Koopmans picture breaks down in the case of inner valence orbitals and the orbital relaxation becomes an important factor in determining its accuracy. Table 5 presents the statistical analysis of IP values for inner valence orbitals as compared to EOMIP-CCSDT values. 20 IP states have been considered in the statistical analysis. It can be seen that the accuracy of the IP values in EOM-CCSD method considerably deteriorates for the inner valence states, with a Max. Abs. Error of 0.73 eV, MAE of 0.44 eV and RMSD of 0.47 eV. The results generally tend to be overestimated in EOM-CCSD, with the minimum spread of error among all the approximate methods. The EOM-CCSD(2) method tends to overestimate the IP values as compared to EOMIP-CCSDT results with a Max. Abs. Error of 1.17 eV, MAE of 0.55 eV and RMSD of 0.63 eV. The results in P-EOM-CCSD(2) considerably deteriorated than that in EOMIP-CCSD, with Max. Abs. Error of 2.49 eV, MAE of 0.63 eV and RMSD of 0.86 eV. The truncated doubles –doubles block does not provide sufficient relaxation necessary for inner valence IP. Consequently, the accuracy of P-EOMIP-CCSD(2) results deteriorates than in outer valence IP, where relaxation has less importance (see Figure 5). The results in P-EOM-CCSD(2) shows considerable spread. Both EOM-CC2 and EOM-CCSD(2)* shows a smaller MAE of 0.36 eV and 0.41 eV, which is less than that observed in EOM-CCSD. However, some inner valence states get very poorly described in both the methods resulting in large error bars, which gets reflected in their higher RMSD values and in STDEV values. For example, $2\sigma_g$ state in $N_2$ shows a drastically high error in EOMIP-CCSD(2)* method. It is also surprising that the maximum error in EOMIP-CCSD(2)* method for that particular state can be higher than that observed in original EOM-CCSD(2), upon which the EOMIP-CCSD(2)* method was designed to improve upon. However, it should be kept in mind that the energy correction in EOMIP-CCSD(2)* method is perturbative, which is less reliable when a state becomes dominated by double excitation, which is the case with $2\sigma_g$ states in $N_2$.

### 3.3 Core Ionization Spectra

The large relaxation effect, associated with the ionization of electron from core orbitals makes the computation of the core-ionization spectra a challenging task in standard ab-initio methods. This is particularly problematic in the EOM-CC method, where the linear excitation operator in the singles and doubles approximation is inadequate to provide the high amount of orbital relaxation effect necessary for a proper description of ionization from core orbital. Table 6 presents the statistical analysis of error for 12 core ionized states as compared to EOM-CCSDT reference values. Many of the core ionized states were difficult to converge in the EOM-CCSDT method even with the small cc-pCVDZ basis set. This has left us with a smaller amount of data in a smaller basis set. The convergence issue in EOM-CC for ionization or excitation from core orbitals is well known[17] and needs special techniques[39] for the solution of that kind of states, which is presently not implemented in any EOM-CCSDT code we have access to.

The EOM-CCSD method shows very poor performance in terms of absolute error and IP values are grossly overestimated, with Max. Abs. error of 2.49 eV, MAE of 1.67 eV and RMSD of 1.73 eV. However, one needs to keep in mind that in terms of relative error, the results are less that 5% of than the benchmark values.

The error becomes worse in EOM-CCSD(2) method, the IP values get severely overestimated and shows Max. Abs. error of 3.85 eV, MAE of 2.12 eV and RMSD of 2.22 eV. The truncated doubles-doubles block in P-EOM-CCSD(2) leads to further reduction of relaxation, resulting in a large increase in error, with Max. Abs. error of 3.84 eV, MAE of 2.60 eV and RMSD of 2.71 eV. It is interesting to note that the IP-CISD method which gives significantly inaccurate results for valence IPs, gives results comparable to that of the more sophisticated EOM-CCSD, with Max. Abs. error of 2.28 eV, MAE of 1.69 eV and RMSD of 1.73 eV. It can be seen from the Figure 6, that EOM-CCSD method systematically overestimates and IP-CISD systematically underestimate the IP values as compared to the EOM-CCSDT values. The spread in the IP-CISD is, however, less than that EOM-CCSD.

Similarly, the EOM-CC2 method also shows much improvement over EOM-CCSD and IP-CISD, with Max. Abs. error of 1.97 eV, MAE of 1.55 eV and RMSD of 1.59 eV, which is even better than the IP-CISD results. However, in terms of the spread of error, it shows quite similar behavior as IP-CISD. The EOM-CCSD(2)* gives the best agreement with EOM-CCSDT, with Max. Abs. Error of 1.19 eV, MAE of 0.51 eV and RMSD of 0.62 eV. The perturbative triples correction tends to correct for the lack of relaxation in EOM-CCSD(2)* due to missing singles amplitudes. It also highlights the importance of relaxation in determining the accuracies of core-ionized states.

# 4 CONCLUSION

In this paper, we have studied the accuracies of different lower scaling variants of EOM-CCSD compared to EOM-CCSDT and CCSD(T) reference. It can be seen that the standard EOM-CCSD method gives the best performance for valence IP. Its performance is comparable to Δ CCSD method for ionization from HOMO. The accuracy of EOM-CCSD method deteriorates for inner valence states compared to that in ionization from outer valence orbitals. The EOMIP-CCSD method shows inferior performance for core-ionized states, where the singles doubles approximation is inadequate to incorporate the orbital relaxation accompanying ionization of core orbitals. Among the approximate variants, the EOM-CCSD(2) method constantly overestimated the IP values for all kinds of ionized states. The introduction of partitioning of the doubles-doubles block does not increase the accuracy of IP values as compared to EOM-CCSD(2) method. Instead, the spread of the error increases in P-EOM-CCSD(2). The EOM-CC2 method shows very poor performance for valence ionized states, in contrast to what observed for valence excited states where the CC2 approximation is remarkably successful. The IP-CISD also fails disastrously for valence IP states with MAE higher than 2 eV for both inner and outer valence IP. However, both EOMIP-CC2 and IP-CISD method gives a surprisingly good performance for core IP, where the results are even better than the standard EOM-CCSD method. Among all approximate methods, the EOMIP-CCSD(2)* method gives comparable performance to EOM-CCSD methods for valence IP and performs even better than the EOM-CCSD method for core IP. However, it tends to show so drastically high error for some of the outlier states.

Therefore, none of the perturbative approximation to EOM-CCSD version gives a balanced description of outer valence, inner valence, and core ionization. The EOMIP-CCSD(2)* looks most promising, among the all the variants tested in this work, However, a lot of ground is still left to cover in the development of lower scaling EOMCC method for ionized states.

## 5. Acknowledgement

The authors acknowledge the grant from CSIR XII[th] five year plan project on Multi-scale Simulations of Material (MSM) and facilities of the Centre of Excellence in Scientific Computing at NCL. A.K.D thanks the Council of Scientific and Industrial Research (CSIR) for a Senior Research Fellowship. S.P. acknowledges the DST J. C. Bose Fellowship project and CSIR SSB grant towards completion of the work.

<u>Supporting Information</u>

Cartesian coordinates of the studied molecules and IP values for the benchmark molecules in different EOMCC methods are given in the supporting information.

*Table 1 : Scaling and storage requirements of various approximate forms of EOM-CCSD*

| Method | Scaling | 3 particle intermidate | 4 particle intermediate |
|---|---|---|---|
| EOM-CCSD | interative O($N^6$) | ☐ | ☐ |
| EOM-CCSD(2) | interative O($N^5$) | ☐ | ☐ |
| P-EOM-CCSD(2) | interative O($N^5$) | ☐ | ☐ |
| EOM-CCSD(2)* | non-interative O($N^6$) | ☐ | ☐ |
| IP-CISD | interative O($N^5$) | ☐ | ☐ |
| EOM-CC2 | interative O($N^5$) | ☐ | ☐ |

*Table 2: Analysis of error for ionization from HOMO in cc-pVTZ basis set compared to CCSD(T) reference*

| Statistical Quantity | EOM-CCSD | EOM-CCSD(2) | IP-CISD | P-EOM-CCSD(2) | EOM-CC(2) | EOM-CCSD(2)* | CCSD |
|---|---|---|---|---|---|---|---|
| Max. Abs. Error | 0.26 | 0.44 | 3,85 | 0.55 | 1.13 | 0.21 | 0.29 |
| MAE | 0.09 | 0.20 | 2.61 | 0.22 | 0.43 | 0.12 | 0.13 |
| RMSD | 0.11 | 0.26 | 2.66 | 0.28 | 0.51 | 0.13 | 0.15 |
| MSD | 0.02 | 0.15 | -2.61 | -0.01 | -0.40 | -0.11 | -0.04 |
| STDEV | 0.11 | 0.22 | 0.54 | 0.29 | 0.33 | 0.07 | 0.14 |

*Table 3: Analysis of error for ionization from HOMO in cc-pVTZ basis set compared to EOM-CCSDT reference*

| Statistical Quantity | EOM-CCSD | EOM-CCSD(2) | IP-CISD | P-EOM-CCSD(2) | EOM-CC(2) | EOM-CCSD(2)* | CCSD |
|---|---|---|---|---|---|---|---|
| Max. Abs. Error | 0.25 | 0.49 | 3,93 | 0.54 | 1.02 | 0.23 | 0.37 |
| MAE | 0.08 | 0.19 | 2.59 | 0.21 | 0.40 | 0.10 | 0.13 |
| RMSD | 0.10 | 0.26 | 2.64 | 0.25 | 0.48 | 0.12 | 0.17 |
| MSD | 0.04 | 0.17 | -2.59 | 0.01 | -0.38 | -0.09 | -0.02 |
| STDEV | 0.09 | 0.20 | 0.53 | 0.26 | 0.30 | 0.08 | 0.17 |

*Table 4: Analysis of error for ionization from outer valence orbitals in cc-pVTZ basis set compared to EOM-CCSDT reference*

| Statistical Quantity | EOM-CCSD | EOM-CCSD(2) | IP-CISD | P-EOM-CCSD(2) | EOM-CC(2) | EOM-CCSD(2)* |
|---|---|---|---|---|---|---|
| Max. Abs. Error | 0.59 | 0.86 | 3.97 | 0.65 | 1.03 | 0.87 |
| MAE | 0.11 | 0.21 | 2.60 | 0.23 | 0.44 | 0.11 |
| RMSD | 0.15 | 0.29 | 2.65 | 0.29 | 0.52 | 0.16 |
| MSD | 0.09 | 0.20 | -2.60 | 0.04 | -0.40 | -0.08 |
| STDEV | 0.13 | 0.21 | 0.52 | 0.29 | 0.32 | 0.14 |

*Table 5: Analysis of error for ionization from Inner valence orbitals in cc-pVTZ basis set compared to EOM-CCSDT reference*

| Statistical Quantity | EOM-CCSD | EOM-CCSD(2) | IP-CISD | P-EOM-CCSD(2) | EOM-CC(2) | EOM-CCSD(2)* |
|---|---|---|---|---|---|---|
| Max. Abs. Error | 0.79 | 1.17 | 3.40 | 2.49 | 1.58 | 2.11 |
| MAE | 0.43 | 0.55 | 2.29 | 0.63 | 0.36 | 0.41 |
| RMSD | 0.48 | 0.63 | 2.45 | 0.86 | 0.54 | 0.75 |
| MSD | 0.29 | 0.44 | -2.29 | 0.31 | -0.32 | -0.30 |
| STDEV | 0.39 | 0.51 | 0.88 | 0.83 | 0.45 | 0.70 |

*Table 6: Analysis of error for ionization from core orbitals in cc-pCVDZ basis set compared to EOM-CCSDT reference*

| Statistical Quantity | EOM-CCSD | EOM-CCSD(2) | IP-CISD | P-EOM-CCSD(2) | EOM-CC(2) | EOM-CCSD(2)* |
|---|---|---|---|---|---|---|
| Max. Abs. Error | 2.52 | 3.85 | 2.28 | 3.84 | 1.97 | 1.19 |
| MAE | 1.67 | 2.12 | 1.69 | 2.60 | 1.55 | 0.51 |
| RMSD | 1.73 | 2.23 | 1.73 | 2.71 | 1.59 | 0.62 |
| MSD | 1.67 | 2.12 | -1.69 | 2.60 | -1.55 | 0.51 |
| STDEV | 0.48 | 0.74 | 0.36 | 0.80 | 0.37 | 0.37 |

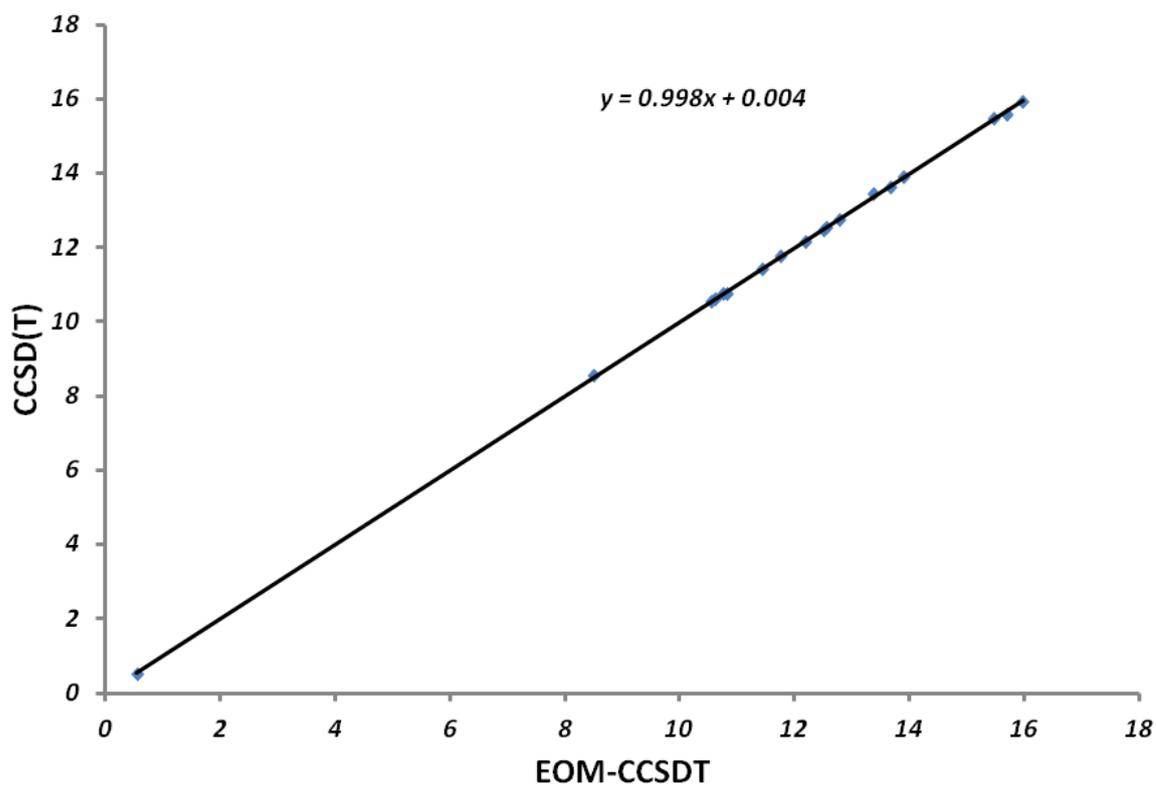

*Figure 1: Plot of IP in CCSD(T) vs EOM-CCSDT for ionization from HOMO in cc-pVTZ basis set*

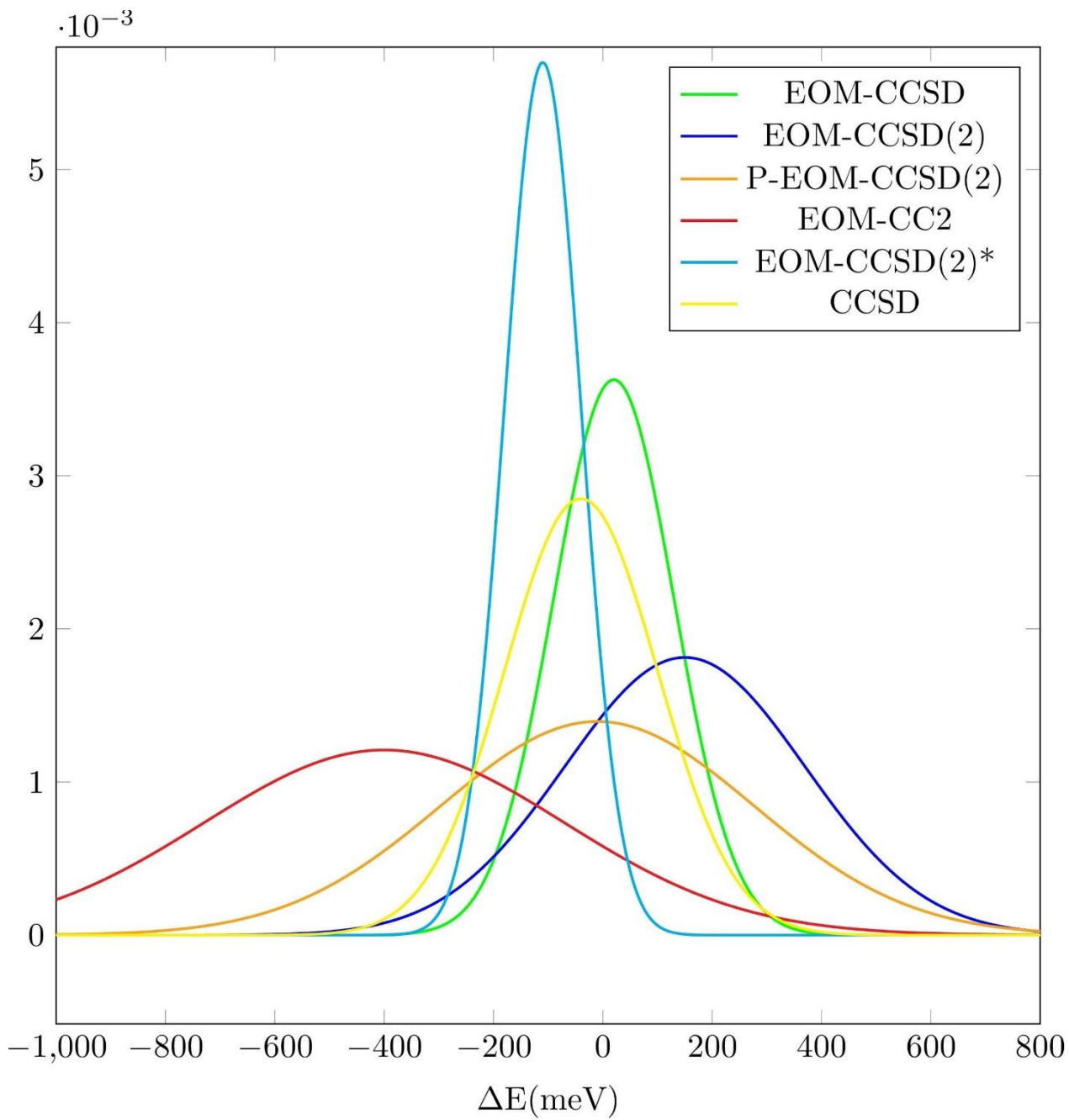

*Figure 2: distribution for ionization from HOMO in cc-pVTZ basis set compared to CCSD(T) reference*

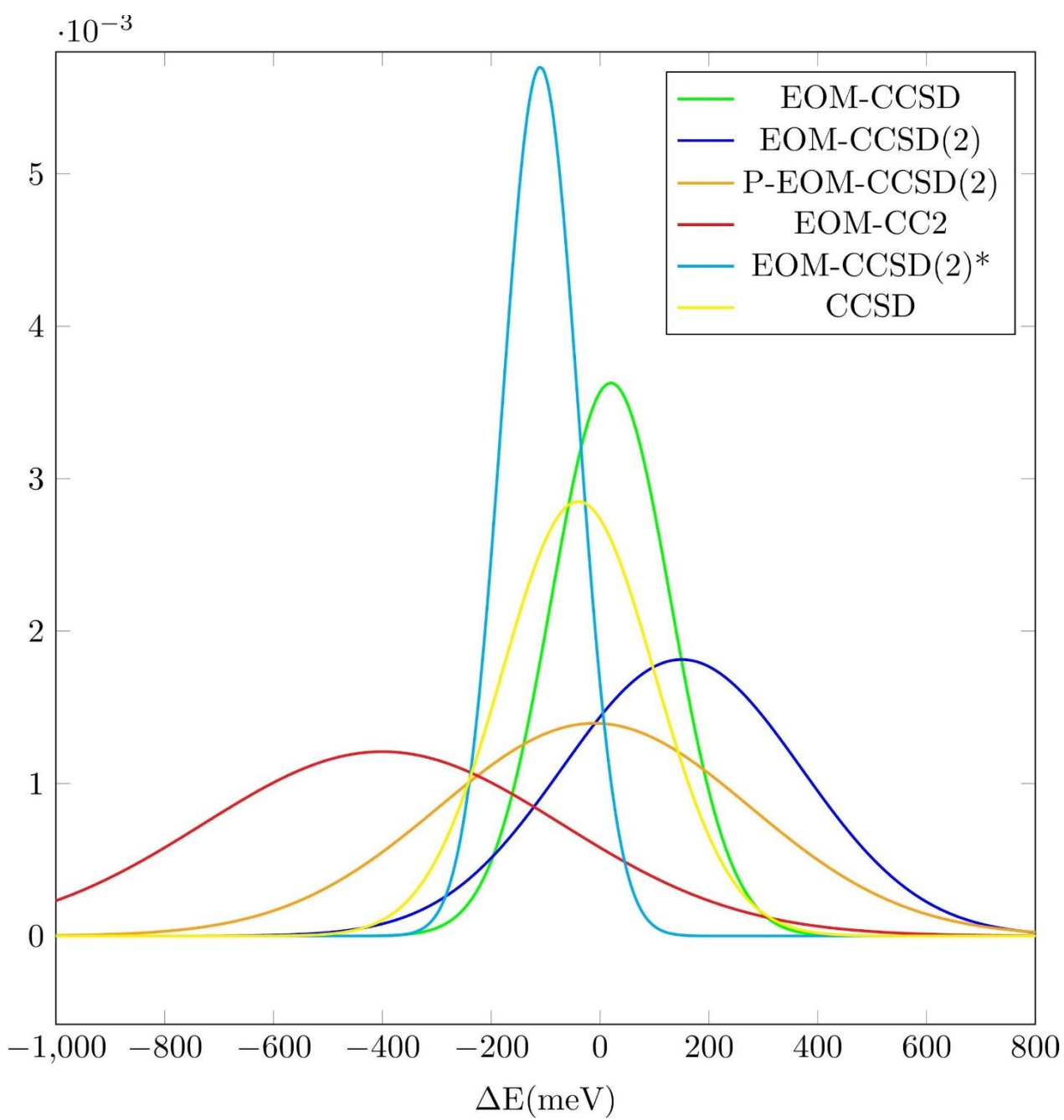

*Figure 3: Error distribution for ionization from HOMO in cc-pVTZ basis set compared to EOM-CCSDT reference*

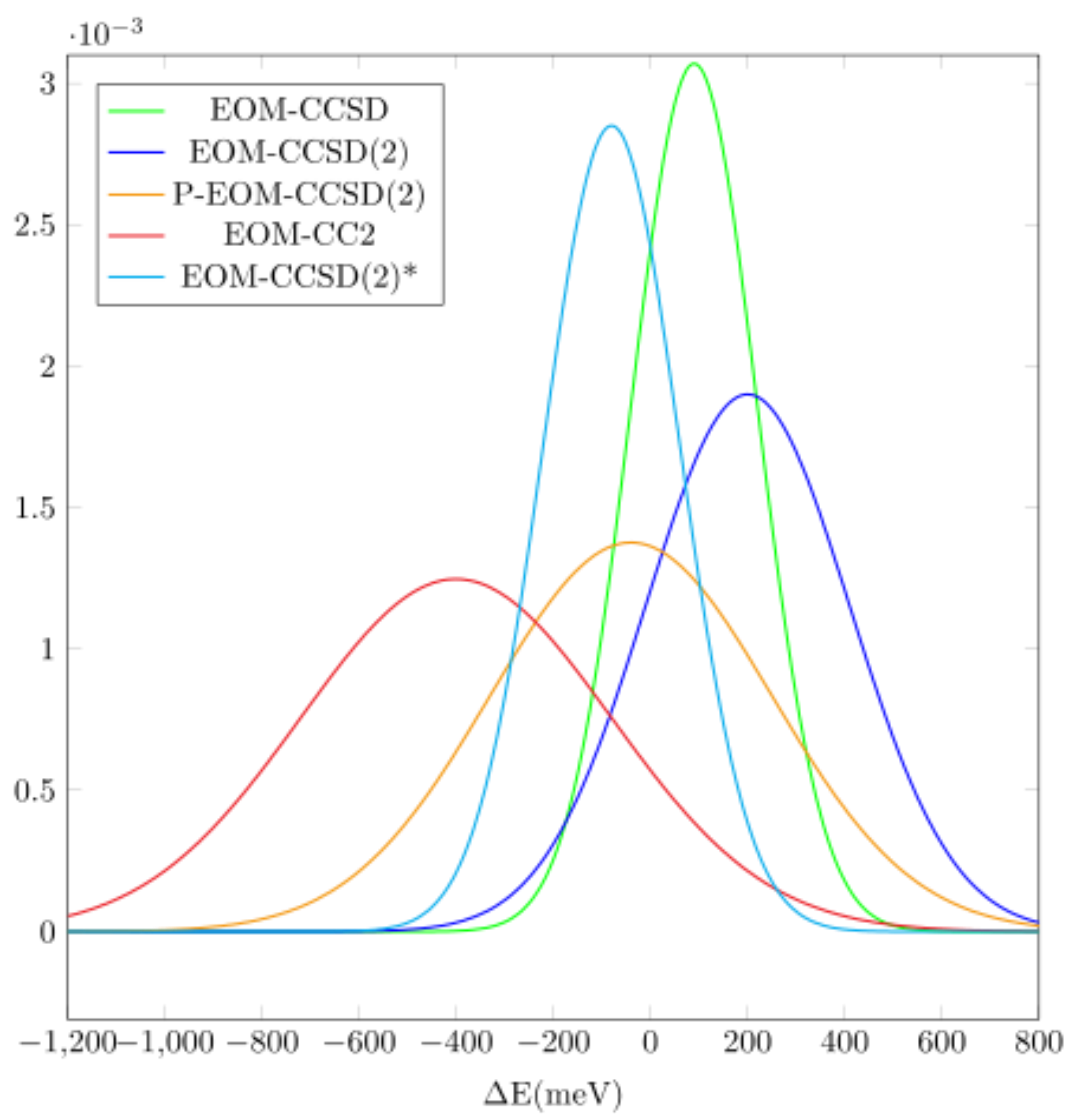

*Figure 4: Error distribution for ionization from outer valence orbital in cc-pVTZ basis set compared to EOM-CCSDT reference*

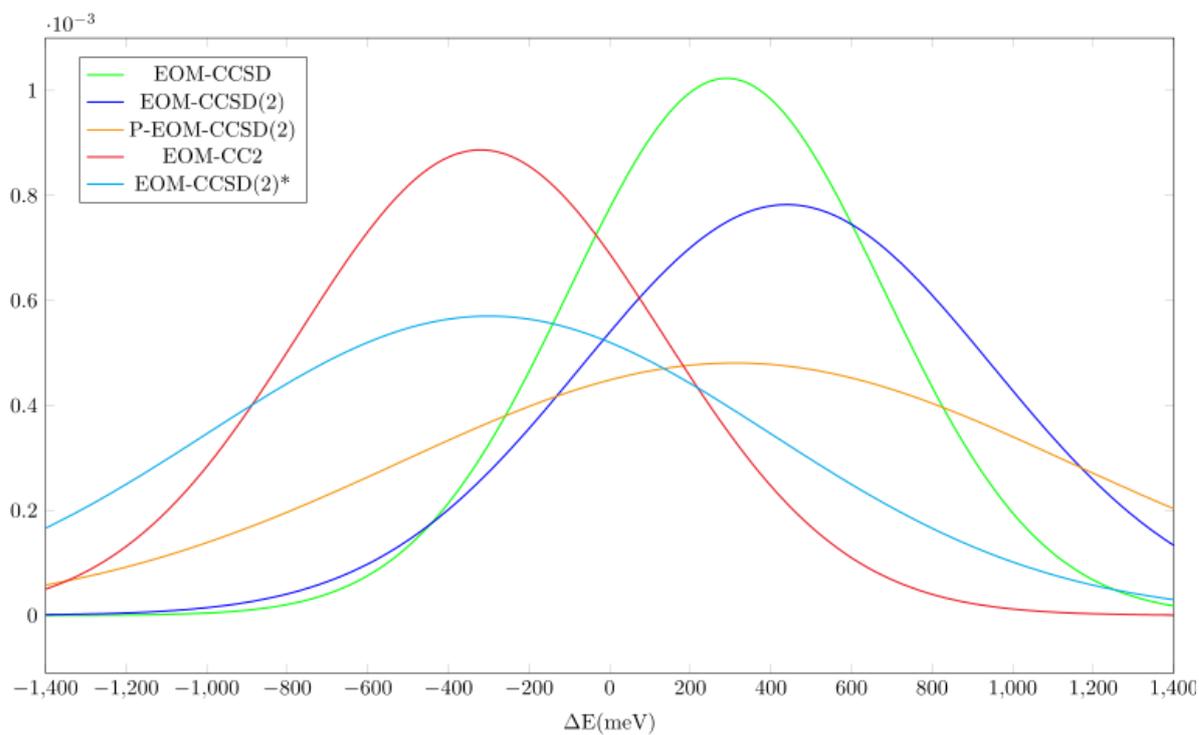

*Figure 5: Error distribution for ionization from inner valence orbital in cc-pVTZ basis set compared to EOM-CCSDT reference*

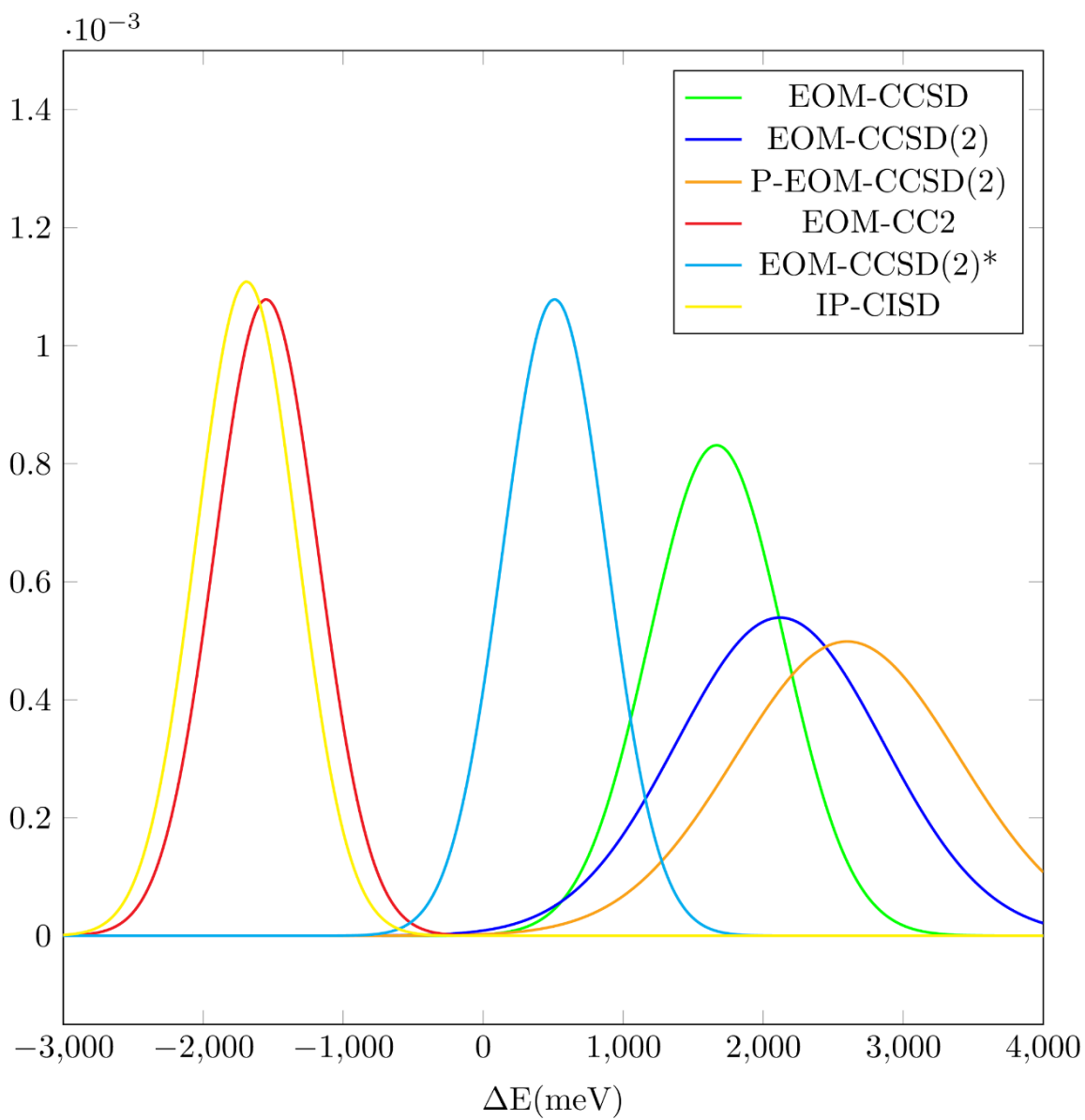

Figure 6: Error distribution for ionization from core orbital in cc-pVTZ basis set compared to EOM-CCSDT reference